\def\[{$$}
\def\]{$$}
\input epsf.sty
\magnification1040
\rightline{FTUAM 06-16}
\rightline{November, 2006}
\smallskip

\centerline{{\bf The Olsson sum rule and the rho Regge pole}}
\bigskip
\bigskip
\noindent{\sl F. J. Yndurain}

\noindent{Departamento de F\'{\i}sica Te\'orica, C-XI}

\noindent Universidad Aut\'onoma de Madrid

\noindent Canto Blanco, Madrid-28049, SPAIN

\vskip1truecm

\noindent
{\bf Abstract}.\quad {\sl We consider the Olsson sum rule, i.e.,
 the forward dispersion relation for pion-pion
scattering  with exchange of isospin unity at threshold. 
We show that, if using the S0, S2 and P wave
expressions  of Colangelo, Gasser and Leutwyler, then either the sum rule is not satisfied or, if 
adjusting the residue of the rho exchange Regge amplitude to have the sum rule satisfied 
(as recently proposed by Caprini, Colangelo and Leutwyler) then the subsequent high energy 
amplitude is in disagreement with experimental pi-pi cross sections.}

\vskip1truecm
\noindent
In a paper written some time ago by Pel\'aez and myself$^{[1]}$ 
(see also ref.~2 for a full discussion of the Regge amplitudes) 
 we remarked that, if in the so-called Olsson
sum rule,
\[2a_0^{(0)}-5a_0^{(2)}=D_{\rm Ol.},\quad
D_{\rm Ol.}\equiv 3M_\pi\int_{4M_\pi^2}^\infty d s\,
{{{\rm Im}\, F^{(I_t=1)}(s,0)}\over{s(s-4M_\pi^2)}}
\]
we use for calculating the quantity ${\rm Im}\, F^{(I_t=1)}(s,0) $ 
the values of the S0, S2 and P wave phase shifts proposed by Colangelo, 
Gasser and Leutwyler $^{[3]}$ at low energy, 
together with experimental information for the other 
waves and for S0, S2 and P waves at intermediate energy; 
and, at high energy, we input the amplitude for rho
exchange deduced in the previously quoted  articles by Pel\'aez and myself from factorization, 
 then the Olsson
sum rule is violated by a bit more than two standard deviations.

Subsequently, Caprini, Colangelo, Gasser and Leutwyler$^{[4]}$
 conceded that we were right,  but stated 
that they were sure that one could find a Regge parametrization that would restore fulfillment of the 
Olsson relation.
 This was done recently, in a paper by
Caprini, Colangelo and  Leutwyler:$^{[5]}$ 
here, these authors  determine
the  parameters for exchange of isospin 1 that make the Olsson sum rule
satisfied, with the $\pi\pi$ amplitudes of Colangelo, Gasser and Leutwyler.$^{[3]}$
 That is to say: they consider the
Olsson sum rule given above, they substitute these low energy amplitudes and 
fix the residue $\beta_\rho$ of the rho trajectory, assumed to give the imaginary part of the amplitude at
high energy,
\[{\rm Im}\,F^{(I_t=1)}(s,0)\simeq \beta_\rho (s/s_0)^{\alpha_\rho},\] 
by requiring the Olsson sum rule to be verified.  

However, the existing 
high energy experimental data can be used to get the 
cross section for exchange of isospin unity.
We can define the cross section $\sigma^{(I_t=1)}(s)$ 
terms of ${\rm Im}\,F^{(I_t=1)}(s,0)$: 
\[\sigma^{(I_t=1)}(s)=
{{4\pi^2}\over{\lambda^{1/2}(s,M^2_\pi,M^2_\pi)}}\,{\rm Im}\,F^{(I_t=1)}(s,0);
\quad \lambda(a,b,c)=a^2+b^2+c^2-2ab-2ac-2bc. \]  The cross section
$\sigma^{(I_t=1)}(s)$  may then be written as
\[\sigma^{(1)}\equiv\sigma^{(I_t=1)}=\sigma_{{\rm tot},\;\pi^+\pi^-}-\sigma_{{\rm tot},\;\pi^-\pi^-},\]
and one can take the $\sigma_{{\rm tot},\;\pi^+\pi^-},\;\sigma_{{\rm tot},\;\pi^-\pi^-}$ from 
experiment.$^{[6]}$ The results are shown in the
accompanying figure, where we plot ({\sl thick line}) against experiment 
what was obtained  by Pel\'aez and Yndur\'ain (loc. cit.) 
either by direct fits to data  
or from factorization; this two methods agree, within the thickness of the line. Then, the  
{\sl dotted line} is  central value of Caprini, Colangelo and  Leutwyler$^{[5]}$ that 
follows from assuming the Olsson sum rule satisfied with the low energy parameters of Golangelo, Gasser and
Leutwyler.

\topinsert{
\setbox1=\vbox{\hsize17.truecm{\epsfxsize=16.truecm\epsfbox{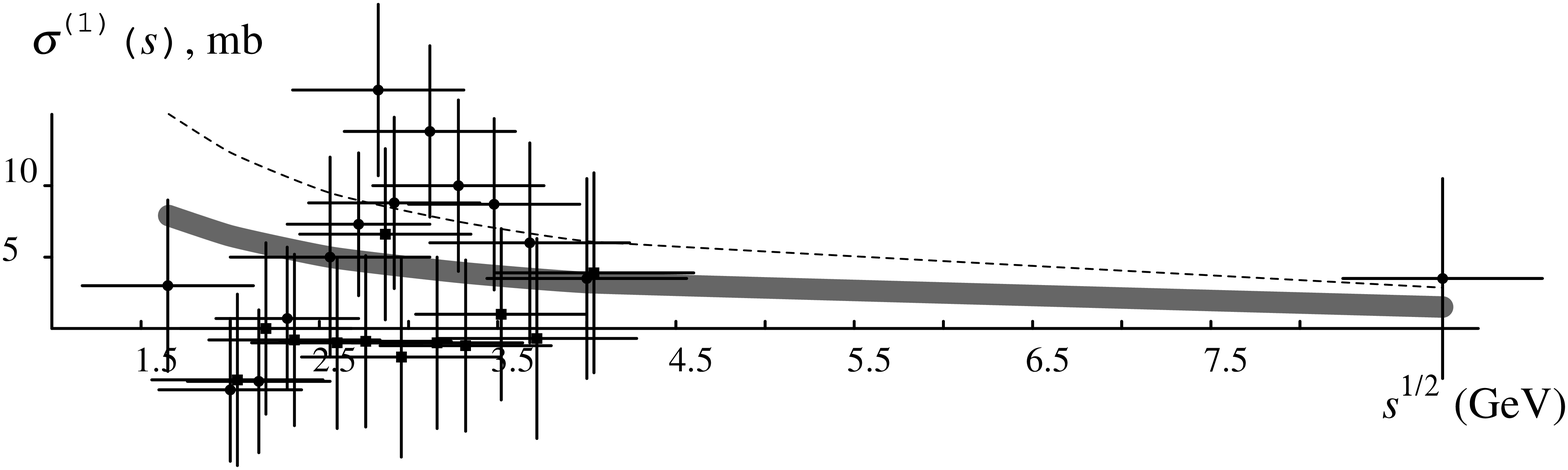}}}
\centerline{{\box1}}
}\endinsert
\bigskip

The data are not very good, but it is clear that the central value that  Caprini, Colangelo and  Leutwyler
get from the Olsson sum rule goes well above what one finds from factorization,
 and also above the majority of
the  experimental data points. One may conclude that
 if  you want a rho Regge amplitude that leads to satisfaction of the 
Olsson sum rule with the Colangelo, Gasser and Leutwyler low energy 
partial waves, such amplitude will disagree with factorization and with most experimental 
information 
for
$\pi\pi$ total cross sections.

\vskip1truecm
\noindent
This note contains only part of the contribution of the author to the  {\sl Quark Confinement and
the Hadron Spectrum} Conference celebrated in the Azores, Portugal.  The rest of the contribution was
identical to that presented at the {\sl IV International Conference on Quarks and 
Nuclear Physics} celebrated in Madrid between 5 and 10 June, 2006,
by Kami\'nski, Pel\'aez and myself.$^{[7]}$.
 
I am grateful to J.~R.~Pel\'aez with whom the essentials of this work were done.
\vskip1truecm
\noindent{References}
\medskip
\noindent
\item{1 }{Pel\'aez, J.~R., and Yndur\'ain, F. J., {\sl Phys. Rev.}
 {\bf D68}, 074005 (2003).}
\item{2 }{Pel\'aez, J.~R., and Yndur\'ain, F. J., {\sl Phys. Rev.} {\bf D69}, 114001
(2004).} 
\item{3 }{Colangelo, G,  
Gasser, J., and Leutwyler, H. {\sl
Nucl. Phys.} {\bf B603},  125, (2001).}
\item{4 }{Caprini, I.,  Colangelo, G.,  Gasser, J.,  and Leutwyler, H. {\sl
Phys. Rev.} {\bf D68}, 074006 (2003).}
\item{5 }{Caprini, I., Colangelo, G., and  Leutwyler, H. {\sl Int. J. Mod. Phys.} {\bf A21}, 954
(2006).}
\item{6 }{Biswas, N. N., et al., {\sl Phys. Rev. Letters}, 
{\bf 18}, 273 (1967) [$\pi^-\pi^-$, $\pi^+\pi^-$ and $\pi^0\pi^-$];
 Cohen, D. et al., {\sl Phys. Rev.}
{\bf D7}, 661  (1973) [$\pi^-\pi^-$];
 Robertson, W. J.,
Walker, W. D., and Davis, J. L., {\sl Phys. Rev.} {\bf D7}, 2554  (1973)  [$\pi^+\pi^-$]; 
Hoogland, W., et al.  {\sl Nucl. Phys.}, {\bf B126}, 109 (1977) [$\pi^-\pi^-$];
Hanlon, J., et al,  {\sl Phys. Rev. Letters}, 
{\bf 37}, 967 (1976) [$\pi^+\pi^-$]; Abramowicz, H., et al. {\sl Nucl. Phys.}, 
{\bf B166}, 62 (1980) [$\pi^+\pi^-$].}
\item{7 }{Kami\'nski, R., Pel\'aez, J.~R., and Yndur\'ain, F. J., hep-ph/0610315.}

\bye